\documentclass[sigconf]{acmart}
\usepackage{courier} 
\usepackage{xcolor}
\usepackage{algpseudocode}
\usepackage{times}
\usepackage{fancyhdr,graphicx,amsmath,amssymb}

\usepackage[linesnumbered,ruled,vlined]{algorithm2e}
\include{pythonlisting} 

\usepackage{soul}
\usepackage{color}
\usepackage{bbding}
\usepackage{amssymb}
\usepackage{makecell}
\usepackage{booktabs}
\usepackage{graphics}
\usepackage{xspace}
\usepackage{wasysym}
\usepackage{amsfonts}
\usepackage{xcolor}
\usepackage{CJKutf8}
\usepackage{multirow}
\usepackage{multicol}
\usepackage{array}
\usepackage{subfig}
\usepackage{amsmath}
\usepackage{float}
\usepackage{hyphenat}
\usepackage{pgfplots}
\usepackage{tikz}
\usepackage[utf8]{inputenc}
\usepackage{newunicodechar}
\usepackage{stackengine,scalerel}
\usepackage{fancyhdr}
\usepackage{graphicx}  %

\AtBeginDocument{%
  \providecommand\BibTeX{{%
    \normalfont B\kern-0.5em{\scshape i\kern-0.25em b}\kern-0.8em\TeX}}}

\newcommand{\system}{MVIN\xspace}

\copyrightyear{2020}
\acmYear{2020}
\setcopyright{acmcopyright}
\acmConference[SIGIR '20] {43rd International ACM SIGIR Conference on Research and Development in Information Retrieval}{July 25--30, 2020}{Virtual Event, China}
\acmBooktitle{43rd International ACM SIGIR Conference on Research and Development in Information Retrieval (SIGIR '20), July 25--30, 2020, Virtual Event, China}
\acmPrice{15.00}
\acmDOI{10.1145/3397271.3401126}
\acmISBN{978-1-4503-8016-4/20/07}

\begin{document}

\fancyhead{}
\fancyfoot{}

\title{MVIN: Learning Multiview Items for Recommendation}

\author{Chang-You Tai}
\affiliation{%
   \institution{Academia Sinica}
   \city{Taipei}
   \country{Taiwan}}
\email{johnnyjana730@gmail.com}

\author{Meng-Ru Wu}
\affiliation{%
   \institution{Academia Sinica}
   \city{Taipei}
   \country{Taiwan}}
\email{ray7102ray7102@gmail.com}

\author{Yun-Wei Chu}
\affiliation{%
   \institution{Academia Sinica}
   \city{Taipei}
   \country{Taiwan}}
\email{yunweichu@gmail.com}

\author{Shao-Yu Chu}
\affiliation{%
   \institution{Academia Sinica}
   \city{Taipei}
   \country{Taiwan}}
\email{shaoyu0966@gmail.com}

\author{Lun-Wei Ku}
\affiliation{%
   \institution{Academia Sinica}
   \city{Taipei}
   \country{Taiwan}}
\email{lwku@iis.sinica.edu.tw}

\begin{abstract}
Researchers have begun to utilize heterogeneous knowledge graphs (KGs) as auxiliary information in recommendation systems to mitigate the cold start and sparsity issues. However, utilizing a graph neural network (GNN) to capture
information in KG and further apply in RS is still problematic as it is unable to see each item's properties from multiple perspectives. To address these
issues, we propose the multi-view item network (\system), a GNN-based recommendation 
model which provides superior recommendations by describing
items from a unique mixed view from user and entity angles. \system learns
item representations from both the user view and the entity view.
From the user view, user-oriented modules score and aggregate features to make
recommendations from a personalized perspective constructed according to KG entities which incorporates user click information. From the entity view, the mixing layer contrasts layer-wise GCN information to further obtain comprehensive features from internal entity-entity interactions in the KG. We evaluate \system on three real-world datasets: MovieLens-1M (ML-1M), LFM-1b
2015 (LFM-1b), and Amazon-Book (AZ-book). Results show that \system
significantly outperforms state-of-the-art methods on these three datasets. In
addition, from user-view cases, we find that \system indeed captures
entities that attract users. Figures further illustrate that mixing layers in a
heterogeneous KG plays a vital role in neighborhood information aggregation.




\end{abstract}

\vspace{-0.5pc}

\ccsdesc{Information systems~\textbf{Recommender systems}}

\keywords{Recommendation, Graph Neural Network, Higher-order Connectivity, Embedding Propagation, Knowledge Graph}

\maketitle

\vspace{-0.5pc}

\section{Introduction}
Recommendation systems (RSs), like many other practical applications with extensive
learning data, have benefited greatly from deep neural networks. Collaborative
filtering (CF) with matrix factorization~\cite{koren2009matrix} is arguably one
of the most successful methods for recommendation in various commercial
fields~\cite{oh2014personalized}. However, CF-based methods' reliance on past
interaction between users and items leads to the cold-start
problem~\cite{schein2002methods}, in which items with no interaction are never
recommended. To mitigate this, researchers have experimented with incorporating auxiliary 
information such as social networks~\cite{Jamali:2010:MFT:1864708.1864736},
images~\cite{Zhang:2016:CKB:2939672.2939673}, and
reviews~\cite{DBLP:journals/corr/ZhengNY17}. 




Among the many types of auxiliary information, knowledge
graphs\footnote{A knowledge graph is typically described as consisting of
entity-relation-entity triplets, where the entity can be an item or an
attribute.}, denoted as KGs hereafter, have widely been used since they can include
rich information in the form of machine-readable entity-relation-entity
triplets. 
Researchers have successively utilized KGs in applications such as node
classification~\cite{hamilton2017inductive}, sentence
completion~\cite{koncel-kedziorski-etal-2019-text}, and summary
generation~\cite{li-etal-2019-coherent}. In view of the success of KGs
in a wide variety of tasks, researchers have developed KG-aware recommendation
models, many of which have benefited from
graph neural networks (GNNs)~\cite{DBLP:journals/corr/abs-1803-03467,DBLP:journals/corr/abs-1904-12575,DBLP:journals/corr/abs-1905-04413,DBLP:journals/corr/abs-1905-07854,DBLP:journals/corr/abs-1904-10322,DBLP:journals/corr/abs-1905-08108,DBLP:journals/corr/abs-1904-12796}
which capture high-order structure in graphs and refine the embeddings of
users and items. For example,
RippleNet~\cite{DBLP:journals/corr/abs-1803-03467} propagates users’ potential
preferences in the KG and explores their hierarchical interests.
Wang et al.~\cite{DBLP:journals/corr/abs-1904-12575} employ an KG graph
convolutional network (GCN)~\cite{2016arXiv160902907K}, which is incorporated in
a GNN to generate high-order item connectivity features.
However, in these models, items look identical to all
users~\cite{DBLP:journals/corr/abs-1803-03467,
DBLP:journals/corr/abs-1905-07854,DBLP:journals/corr/abs-1904-12796}, and using GCN with KGs still has drawbacks such as missing comparisons between entities of different layers~\cite{DBLP:journals/corr/abs-1905-00067}.

%
%
%
%
%

We further give some examples that explain the \textit{user view} and
the \textit{entity view}. Imagine some users are
interested in books of the same author, and other users are interested in a certain
book genre, where authorship and genre are two relations between the book
and its neighborhood (author, genre type) in the knowledge base. We can say
that in the real world every user has a different view of a given item. In the
\textit{entity-view}, item representations are defined by the entities
connected to it in the KG. A sophisticated representation can be generated by
incorporating smart operations of entities.  For example, this paper refines it
by leveraging the layer-wise entity difference to keep information from neighborhood entities. To illustrate the need for
this difference feature, imagine that we seek
to emphasize a \emph{new} actor in a movie directed by a \emph{famous}
director, 
contrasting entities related to the famous director at the second                 
layer to the director at the first layer will have stronger expressiveness than   %
aggregating all of the directors he has co-worked with back him.              %

Overall, there are still challenges with GNN-based recommendation models:
(1) user-view GNN enrichment and (2) entity-view GCN refinement.
In this paper, we investigate GNN-based recommendation and propose
a network that meets the above challenges.
We propose a knowledge graph multi-view item network (\system), 
a GNN-based recommendation model equipped with user-entity and entity-entity
interaction modules. To enrich user-entity interaction, we first learn the
KG-enhanced user representations, using which the user-oriented modules 
characterize the importance of relations and informativeness for each entity.
To refine the entity-entity interaction, we propose a mixing layer to further
improve embeddings of entities aggregated by GCN and allow \system to capture
the mixed GCN information from the various layer-wise neighborhood features.
Furthermore, to maintain computational efficiency and approach the panoramic view of the whole neighborhood, we adopt a stage-wise strategy~\cite{pmlr-v44-Barshan2015} and sampling
strategy~\cite{DBLP:journals/corr/abs-1904-12575,DBLP:journals/corr/abs-1806-01973} to better utilize KG information.
%
%
%
%
%
%

%
%
We evaluate \system performance on three real-world datasets: ML-1M, LFM-1b,
and AZ-book. For click-through rate (CTR) prediction and top-$N$ recommendation,
\system significantly outperforms state-of-the-art models.
Through ablation studies, we further verify the effectiveness of each component
in \system and show that the mixing layer plays a vital role in both homogeneous and
heterogeneous graphs with a large neighborhood sampling size.
%
%
%
Our contributions include: 
\begin{itemize}

\item We enable the user view and personalize the GNN.



\item We refine item embeddings from the entity view by a wide and deep GCN which
brings in layer-wise differences to high-order connectivity. 



\item We conduct experiments on three real-world datasets with KGs of different
sizes to show the robustness and superiority of \system.\footnote{We release the codes and datasets at \href{https://github.com/johnnyjana730/MVIN/}{https://github.com/johnnyjana730/MVIN/}} In addition, we
demonstrate that \system captures entities which identify user interests, and
that layer-wise differences are vital with large neighborhood sampling sizes
in heterogeneous KGs.

\end{itemize}



\vspace{-1.0 pc}

\section{Related Work}

For recommendation, there are other models that leverage KGs, and there are other models
that consider interaction between users and items. We introduce these below.

\subsection{KG-aware Recommendation Models}
In addition to graph neural network (GNN) based
methods, there are two other categories of KG-aware recommendation.

The first is embedding-based
methods~\cite{10.1145/3308558.3313705,Huang:2018:ISR:3209978.3210017_emb,Zhang:2016:CKB:2939672.2939673_emb,Wang:2018:DDK:3178876.3186175_emb, DBLP:journals/corr/abs-1902-06236, Zhang2018LearningOK},
which combine entities and relations of a KG into continuous vector spaces and
then aid the recommendation system by enhancing the semantic representation.
For example, DKN~\cite{2018arXiv180108284W} fuses semantic-level and
knowledge-level representations of news and incorporates KG representation into
news recommendation. In addition, CKE~\cite{Zhang:2016:CKB:2939672.2939673}
combines a CF module with structural, textual, and visual knowledge into a
unified recommendation framework. However, these embedding-based 
knowledge graph embedding (KGE) algorithms methods are more suitable for
in-graph applications such as link prediction or KG completion rather than for
recommendation~\cite{8047276}. Nevertheless, we still 
select~\cite{Zhang:2016:CKB:2939672.2939673} for comparison.

The second category is path-based
methods~\cite{Hu:2018:LMB:3219819.3219965_path,Sun:2018:RKG:3240323.3240361_path,DBLP:journals/corr/abs-1811-04540_path,Yu:2014:PER:2556195.2556259_path,Zhao:2017:MBR:3097983.3098063_path},
which utilize meta paths and related user-item pairs, exploring patterns of
connections among items in a KG. For instance,
MCRec~\cite{Hu:2018:LMB:3219819.3219965_path} learns an explicit representation
for meta paths in recommendation. In addition, it considers the mutual
effect between the meta path and user-item pairs. Compared to
embedding-based methods, path-based methods use the graph algorithm directly to
exploit the KG structure more naturally and intuitively. However, they rely
heavily on meta paths, which require domain knowledge and manual labor to
process, and are therefore poorly suited to end-to-end
training~\cite{DBLP:journals/corr/abs-1803-03467}. We also provide the
performance of the state-of-the-art path-based
model~\cite{Hu:2018:LMB:3219819.3219965_path} as a baseline for comparison.

\subsection{User-Item Interaction}

As users and items are two major entities involved in recommendation,
many works attempt to improve recommendation performance by studying
user-item interaction.

For example, as a KG-aware recommendation model, Wang et
al.~\cite{DBLP:journals/corr/abs-1904-12575} propose KGCN, which characterizes
the importance of the relationship to the user. However, the aggregation method in
KGCN does not consider the informativeness of entities different from the user. Hu
et al.~\cite{MRECinproceedings} propose MCRec, which considers users'
different preferences over the meta paths. Nevertheless, it neglects semantic
differences of relations to users. Also, they do not employ GCN and thus
information on high-order connectivity is limited.

With their KG-free recommendation model, Wu et al.~\cite{2019arXiv190705559W} consider
that as the informativeness of a given word may differ between users, they propose
NPA, which uses the user ID embedding as
the query vector to differentially attend to important words and news according
to user preferences. An et al.~\cite{inproceedingslong_short_term} consider
that users typically have both long-term preferences
and short-term interests, and propose LSTUR which adds user representation
into the GRU to capture the user's individual long- and short-term interests.

\section{Recommendation Task Formulation}

Here we clarify terminology used here
and explicitly formulate \system, the proposed 
GNN-based recommendation model.

In a typical recommendation scenario, the sets of users and items are denoted
as \(\mathcal{U} = \{u_1,u_2...\}\) and \(\mathcal{V} = \{v_1,v_2...\}\), and
the user-item interaction matrix \(\mathbf{Y}=\{y_{uv} \mid u \in \mathcal{U},
v \in \mathcal{V}\}\) is defined according to implicit user feedback.
If there is an observed interaction between user $u$ and item $v$, \(y_{uv}\)
is recorded as \(y_{uv}=1\); otherwise \(y_{uv}=0\).
In addition, to enhance recommendation quality, we leverage the information
in the knowledge graph $\mathcal{G}$, which is comprised of
entity-relation-entity triplets \(\{(h,r,t)|h,t \in \mathcal{E}, r \in
\mathcal{R}\}\).
Triplet $(h,r,t)$ describes relations $r$ from head entity $h$ to
tail entity $t$, and $\mathcal{E}$ and $\mathcal{R}$ denote the set of entities
and relations in $\mathcal{G}$.
Moreover, an item $v \in \mathcal{V}$ may be associated with one or more entities
$e$ in $\mathcal{G}$; $N(v)$ refers to these neighboring entities around $v$.
Given interaction matrix $\mathbf{Y}$ and knowledge graph $\mathcal{G}$, we
seek to predict whether user $u$ has a potential interest in item $v$. The
ultimate goal is to learn the prediction function \(\hat{y}_{uv} =
\mathcal{F}(u,v;\Theta, \mathcal{G})\), where $\hat{y}_{uv}$ is the probability
that user $u$ engages with item $v$, and $\Theta$ stands for the model
parameters of function $\mathcal{F}$.

\section{\system}

\begin{figure*}[htb]
\centering
\includegraphics[scale=0.38, trim={0 0 0 0}]{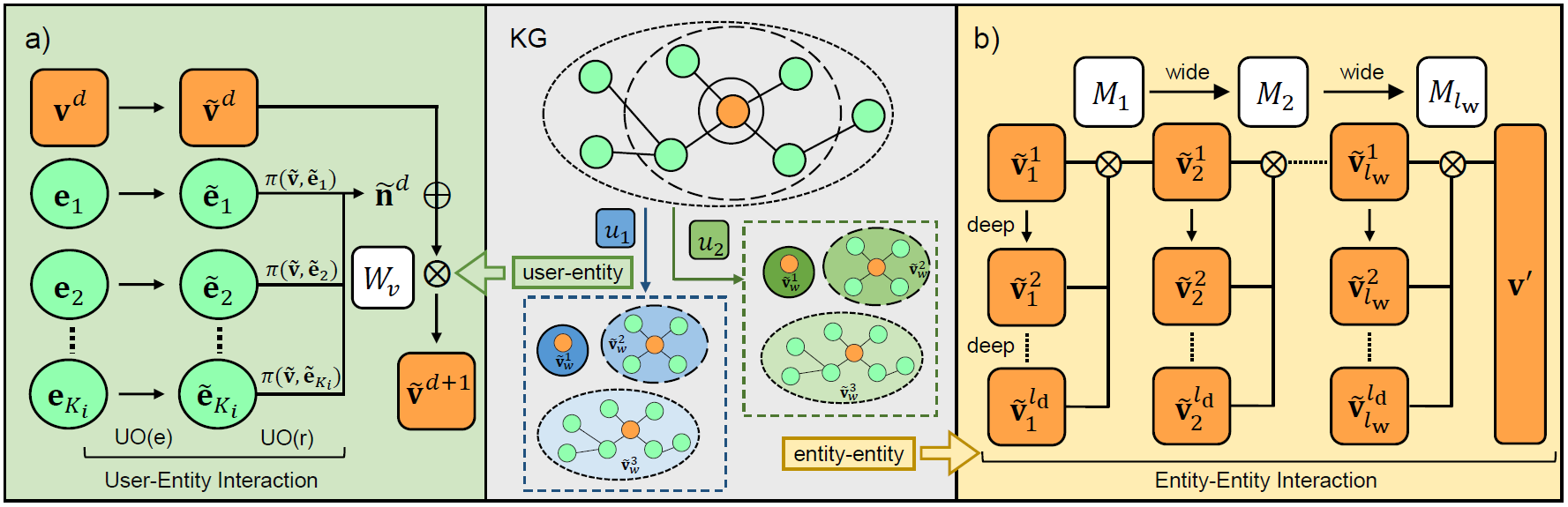}
\vspace{-0.4 cm}
\caption{\system framework, which enhances item
representations through user-entity and entity-entity interaction. For
user-entity interaction, it contains (a) user-oriented relation attention UO(r)
and entity projection modules UO(e) to collect KG entity information from a
user perspective. For entity-entity interaction, the mixing layer allows
\system not only to (a) aggregate high-order connectivity information but also
to (b) mix layer-wise GCN information.}
\label{fig:KGCN_high_orderneighborhoods_information}
\end{figure*}

We describe in detail \system, the proposed recommendation model, shown in
Figure~\ref{fig:KGCN_high_orderneighborhoods_information}, which enhances
item representations through user-entity
interaction, which describes how \system collects KG entities information from a
user perspective, and entity-entity interaction, which helps \system not only to aggregate high-order connectivity information but also to mix layer-wise GCN information.

\subsection{User-Entity Interaction} \label{User-Oriented Mixing Information}

To improve the user-oriented performance, we split user-entity interaction into 
user-oriented relation attention, user-oriented entity projection, and
KG-enhanced user representation.


\subsubsection{User-Oriented Relation Attention} 
\label{User-Oriented relation attention} 
When \system collects information from the neighborhood of the given item in the
KG, it scores each relation around the item in a user-specific way.
The proposed user-oriented relation attention mechanism utilizes the
information of the given user, item, and relations to determine which
neighbor connected to the item is more informative. For instance, some users
may think the film \emph{Iron Man} is famous for its main actor Robert Downey Jr.;
others may think the film \emph{Life of Pi} is famous for its director Ang
Lee. Thus each entity of neighborhood is weighted by dependent scores
${\pi}^{u}_{r_{v,e}}$, where $u$ denotes a different user; ${r_{v, e}}$ denotes
the relation $r$ from entity $v$ to neighboring entity $e$ 
(the formulation of the scoring method is given below). We
aggregate the weighted neighboring entity embeddings and generate the final
user-oriented neighborhood information $\textbf{n}$ as
\begin{align}
   \textbf{n} &= \sum_{e \in \mathcal{N}(v)} \, \widetilde{\pi}^{u}_{r_{v,e}} \textbf{e} \\
   \widetilde{\pi}^{u}_{r_{v,e}} &= \pi(\textbf{v},\textbf{e}) =  \frac{\exp(\pi^{u}_{r_{v,e}})}{\sum_{e' \in \mathcal{N}(v)}\exp(\pi^{u}_{r_{v,e'}})}
\end{align}

To calculate the neighbor's dependent score ${\pi}^{u}_{r_{v,e}}$, we first
concatenate relation $\textbf{r} \in \mathbb{R}^{s}$, item representation
$\textbf{v} \in \mathbb{R}^{s}$, and user embedding $\textbf{u} \in
\mathbb{R}^{s}$, and then transform these to generate the final
user-oriented score ${\pi}^{u}_{r_{v,e}}$ as
\begin{align}
   \pi^{u}_{r_{v,e}} &= W_r(\mathrm{concat}([\textbf{u},\textbf{r},\textbf{v}])) + \textbf{b}_r,
\end{align}
where $W_r \in \mathbb{R}^{3s}$ and $\textbf{b}_r \in \mathbb{R}$ are trainable
parameters.


   
   
\subsubsection{User-Oriented Entity Projection} \label{User-Oriented entity projection} 
To further increase user-entity interaction, we propose a user-oriented entity
projection module. For different users, KG entities should have different
informativeness to characterize their properties. For instance, in a movie
recommendation, the user's impression of actor Will Smith varies from
person to person. Someone may think of him as a comedian due to the film
\emph{Aladdin}, while others may think of him as an action actor due to the film \emph{Bad Boys}.
Therefore, the entity projection mechanism refines the entity embeddings
by projecting each entity $\textbf{e}$ onto user perspective
$\textbf{u}$, where the projecting function can be either linear or non-linear:
\vspace{-0.2 pc}
\begin{align}
     \ \tilde{\textbf{e}} &=  W_e(\textbf{e} + \textbf{u})+ \textbf{b}_e \\
    \tilde{\textbf{e}} &=  \sigma(W_e(\textbf{e} + \textbf{u})+\textbf{b}_e)
\end{align}
where $W_e$ and $\textbf{b}_e$ are trainable parameters and $\sigma$ is the
non-linear activation.


Thus the user-oriented entity projection module can be seen as an early
layer which increases user-entity interactions. Then, the user-oriented
relation attention module aggregates the neighboring information in a
user-specific way.

\subsubsection{KG-Enhanced User Representation} 
\label{KG-Enhanced User Preference} 
\begin{figure}[htb]
\centering
\includegraphics[scale=0.28, trim={0 0 0 0}]{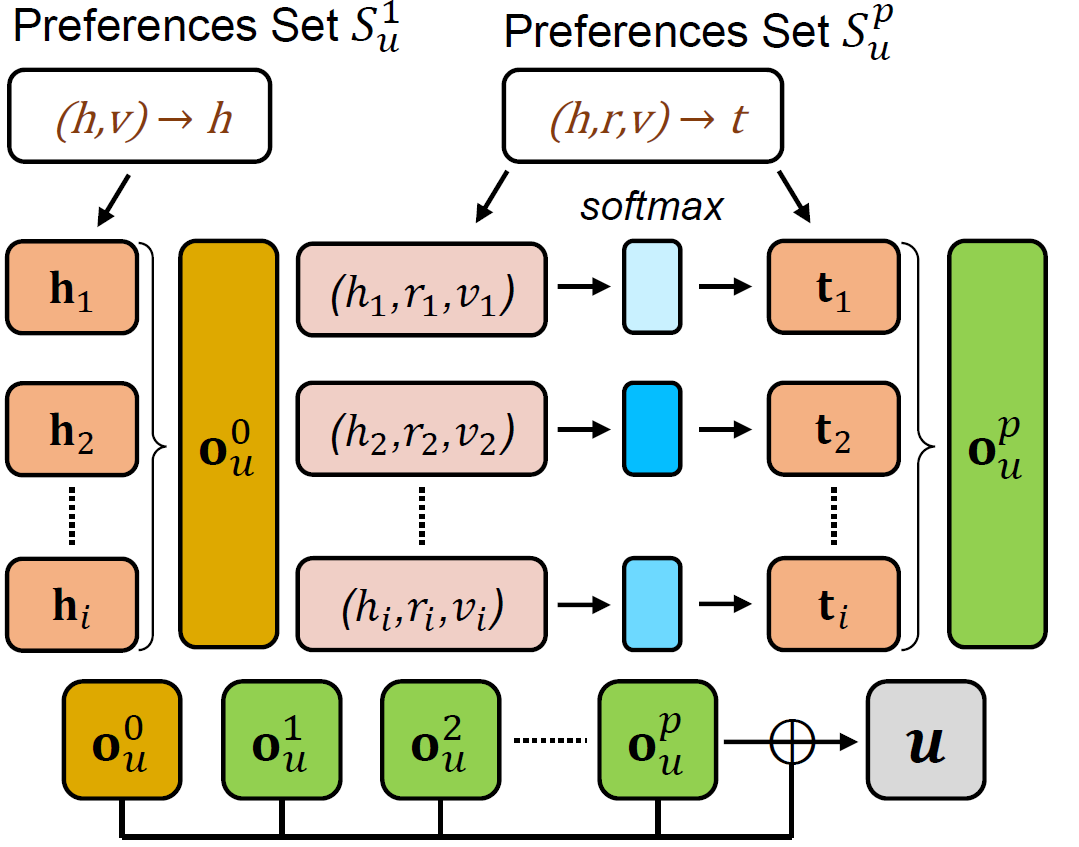}
\caption{KG-enhanced user representation in \system. At hop $p$, user 
preference set $S^{p}_{u}$ is propagated to generate user preference
responses $\textbf{o}^{p}_{u}$, after which all hops of user preference 
responses are integrated to generate the KG-enhanced user representation
\textbf{u}.}
\vspace{-1.5 pc}
\label{fig:KGPH_preference}
\end{figure}

To enhance the quality of user-oriented information received from
previous sections, we enrich user representations constructed according to KG entities which incorporates user click information~\cite{DBLP:journals/corr/abs-1803-03467}. For example, if a user watched \emph{I, Robot}, we find \emph{I, Robot} is
acted by Will Smith, who also acts in \emph{Men in Black} and \emph{The Pursuit of Happyness}.
Capturing user preference information from the KG relies on consulting all relevant
entities in KG and the connections between entities help us to find the potential user interests.
The extraction of user preference also fits the proposed user-oriented modules; in the user's mind, the icon of a famous actor is defined not only by the movies they have watched but also by the movies in the KG that the user is potentially interested in. In our example, if the user has potential interests in Will Smith, the modules would quickly focus on other films he acted.
In sum, the relevant KG entities model the user representation and by KG-enhanced user representation, the user-oriented information is enhanced as well.\footnote{In
Section~\ref{Effect of KG-enhanced User-Orient Information}, this design helps \system to focus on entities which the user may show interest in given the items that the user has interacted with.} The overall process is shown in Figure~\ref{fig:KGPH_preference} and Algorithm 1.


\begin{algorithm}[h]
\label{alg:Propagation_User_Preference}
\small
\DontPrintSemicolon
\SetKwFunction{FMain}{}
\SetKwProg{Fn}{KGUR}{:}{}
\Fn{\FMain{$u$}}{
    \For{$p = 1,...,l_\mathrm{p}$}{
        $\textbf{o}^{p}_{u} \gets \sum_{(h_i,r_i,t_i)\in \mathcal{S}^{p}_{u}}k_{i}\textbf{t}_{i}$; \\
    }
    $ \textbf{o}^{0}_{u} \&= \sum_{h_i\in \mathcal{S}^{1}_{u}} a_{i}\textbf{h}_i$; \\
    $ \textbf{o}_u = \mathrm{concat}([\textbf{o}^{0}_{u}, \textbf{o}^{1}_{u}, \dots, \textbf{o}^{l_\mathrm{p}}_{u}])$;\\
    $\textbf{u} = W_o\textbf{o}_u + \textbf{b}_o$;\\
\textbf{return} $\textbf{u}$;\\
}
\caption{KG-Enhanced User Representation}
\end{algorithm}

\textbf{Preference Set} We first initialize the preference set. For user $u$, the
set of items that the user has interacted with, $\mathcal{V}_u =\lbrace$$v|y_{uv}=1$$\rbrace$, is
treated as the starting point in $\mathcal{G}$, which is then explored along the relations to
construct the preference set $\mathcal{S}_{u}$ as

\vspace{-0.5 pc}

\begin{align} 
  \mathcal{E}^{p}_{u} &= \lbrace t |(h,r,t) \in \mathcal{G} \, \textrm{and} \, h \in \mathcal{E}^{p-1}_{u} \rbrace, \, p= 1,2,...,l_\mathrm{p},
\end{align}

\vspace{-0.5 pc}

where $\mathcal{E}^{0}_{u}$ = $\mathcal{V}_u$; $\mathcal{E}^{p}_{u}$ records the $p$ hop entities linked from entities at previous ${p-1}$ hop.
\vspace{-0.5 pc}

\begin{align} 
  \mathcal{S}^{p}_{u} &= \lbrace (h,r,t)|(h,r,t) \in \mathcal{G} \, \textrm{and} \, h \in \mathcal{E}^{p-1}_{u} \rbrace,\, p= 1,2,...,l_\mathrm{p},
\end{align}

\noindent
where $\mathcal{S}^{p}_{u}$ is the
preference set at hop $p$.
Note that $\mathcal{E}^{p}_{u}$ is only tail entities and $\mathcal{S}^{p}_{u}$
is the set of knowledge triples, $p$ represents the hop(s), and $l_p$ is the number of preference hops.


\textbf{Preference Propagation} 
\label{Propagation method}  
The KG-enhanced user representation is constructed by user preference responses
$\textbf{o}_{u}$ generated by propagating preference set $\mathcal{S}_{u}$. 


First, we define at hop 0 the user preference responses $\textbf{o}^{0}_{i}$ which
is calculated from the user-clicked items $h_{i} \in \mathcal{S}^{1}_{u}$; taking into 
account different items representations $\textbf{v}$ assigns different
degrees of impact to the user preference response:
\begin{align}
   \textbf{o}^{0}_{u} &= \sum_{h_{i} \in \mathcal{S}^{1}_{u}} a_{i}\textbf{h}_{i} \\
    a_{i} &= \mathrm{softmax}_{i}(W_a[\textbf{h}_{i},\textbf{v}]) 
\end{align}
where $W_a$ is a trainable parameter. 

Second, at hop $p$, where $p>0$, user preference responses
$\textbf{o}^{p}_{u}$ are computed as the sum of the tails weighted by the
corresponding relevance probabilities $k_{i}$ as
\begin{align} 
   \textbf{o}^{p}_{u} &= \sum_{(h_{i},r_{i},t_{i})\in {\mathcal{S}}^{p}_{u}}k_{i}\textbf{t}_{i} \,,  p = 1,2,...,l_\mathrm{p}, \\
   k_{i} &= \mathrm{softmax}(\textbf{v}^{T}\textbf{R}_{i}\textbf{h}_{i}) 
\end{align}
where $\textbf{h}_{i} \in \mathbb{R}^{s}$, $\textbf{R}_{i} \in \mathbb{R}^{s
\times s}$, $\textbf{t}_i \in \mathbb{R}^s$, and $\textbf{v} \in
\mathbb{R}^{s}$ are the embeddings of heads $h_i$, relations $r_i$, tails $t_i$,
and item $v$. The relation space embedding $\textbf{R}$ helps to calculate the
relevance of item representation $\textbf{v}$ and entity representation
$\textbf{h}$. 

After integrating all user preference responses $\textbf{o}^{p}_{i}$, we
generate the final preference embedding of user \textbf{u} $\in \mathbb{R}^{s}$
as
\begin{align}
   \textbf{o}_{u} &= \mathrm{concat}([\textbf{o}^{0}_{u}, \textbf{o}^{1}_{u}, \dots, \textbf{o}^{l_\mathrm{p}}_{u}]), \\
   \textbf{u} &= W_o\textbf{o}_{u} + \textbf{b}_o
\end{align}
where $W_o$ and $\textbf{b}_o$ are trainable parameters.


\subsection{Entity-Entity Interaction}

\begin{algorithm}[h]
\label{alg:Integrating_High-order_Neighborhood}
\small
\DontPrintSemicolon
\SetKwFunction{FMain}{}
\SetKwProg{Fn}{MixLayer}{:}{}
\Fn{\FMain{$v,\textbf{u}$}}{
    $ \tilde{\textbf{e}} = \sigma(W_e \cdot (\textbf{e} + \textbf{u}) + \textbf{b}_e)), \forall e \in \mathcal{G};$ \\
    $ \tilde{\textbf{e}}{^{u,1}_{1}} \gets v$; \\
    
    \For{$w = 1,...,{l_{\mathrm{w}}}-1$}{
    \For{$d = 1,...,{l_{\mathrm{d}}}-1$}{
    \For{$e \in \mathcal{G}$}{
        $ \tilde{\textbf{n}}{^{d}_{w}} \gets {\sum_{e' \in \mathcal{N}(e)}} \tilde{\pi}{^{u}_{r_{\tilde{e},\tilde{e}'}}}{\tilde{\textbf{e}}'}{^{d}_{w}};$ \\
        $ \tilde{\textbf{e}}{^{d+1}_{w}} \gets \mathrm{agg}(\tilde{\textbf{e}}{^{d}_{w}},\tilde{\textbf{n}}{^{d}_{w}})$; \\
    }}
    $\tilde{\textbf{e}}{^{1}_{w+1}} = M_w(\mathrm{concat}([\tilde{\textbf{e}}{^{1}_{w}}, \tilde{\textbf{e}}{^{2}_{w}}, \dots, \tilde{\textbf{e}}{^{l_\mathrm{d}}_{w}}]))
    $;}
\textbf{return} $\tilde{\textbf{e}}{^{1}_{l_{\mathrm{w}}}}$;}
\caption{Layer Mixing}
\end{algorithm}

In entity-entity interaction, we propose layer mixing and focus on 
capturing high-order connectivity and mixing layer-wise
information. We introduce these two aspects in terms of depth and width,
respectively; the overall process, combined with the method mentioned in Section~\ref{User-Oriented Mixing Information}, is shown in Figure~\ref{fig:KGCN_high_orderneighborhoods_information} and Algorithm 2.

For depth, we integrate user-oriented information obtained as described in
Section~\ref{User-Oriented Mixing Information}, yielding high-order
connectivity information to generate entity $\tilde{\textbf{v}}^{d}_{w}$ and
neighborhood information $\tilde{\textbf{n}}^{d}_{w}$, followed by aggregation as
$\mathrm{agg}(\cdot)$: $\mathbb{R}^{s} \times \mathbb{R}^{s} \to
\mathbb{R}^{s}$ to generate the next-order representation
$\tilde{\textbf{v}}^{d+1}_{w}$.

leveraging the layer-wise entity difference
For width, to allow comparisons between entities of different order~\cite{DBLP:journals/corr/abs-1905-00067}, we mix the feature representations of neighbors at various
distances to further improve the performance of subsequent 
recommendation.\footnote{In
Section~\ref{Effect_of_mixing_layer-wise_GCN_information_high_sampling_size},
this is shown to help \system to improve results given large neighbor sampling
sizes.} Specifically, at each layer, we utilize layer matrix $M_w$ to mix layer-wise
GCN information ($\tilde{\textbf{v}}^{1}_{w}$,
$\tilde{\textbf{v}}^{2}_{w}$,...,$\tilde{\textbf{v}}^{d}_{w}$) and generate
the next wide layer entity representation $\tilde{\textbf{v}}^{1}_{w+1}$ as
\setlength{\abovedisplayskip}{3pt}
\setlength{\belowdisplayskip}{3pt}
\begin{align}
   \tilde{\textbf{v}}^{1}_{w+1} &= M_w(\mathrm{concat}([\tilde{\textbf{v}}^{1}_{w}, \tilde{\textbf{v}}^{2}_{w}, \dots, \tilde{\textbf{v}}^{l_\mathrm{d}}_{w}])) \\
\tilde{\textbf{v}}^{d+1}_{w}&= \mathrm{agg}(\tilde{\textbf{v}}^{d}_{w}, \tilde{\textbf{n}}^{d}_{w}) =\sigma(W_v(\tilde{\textbf{v}}^{d}_w+\tilde{\textbf{n}}^{d}_{w})+\mathbf{b}_v)
\end{align}
where $w = 1,...,l_\mathrm{w}-1$, $d$=1,2,...,$l_\mathrm{d}$-1; $l_\mathrm{w}$
and $l_\mathrm{d}$ are the number of wide and deep hops, respectively; $W_v$ and $\mathbf{b}_v$ are trainable
parameters.

\vspace{-1.0 pc}

\vspace{1.0 pc}

\subsection{Learning Algorithm}


The formal description of the above training step is presented in
Algorithm 3. For a given user-item pair
($u$,$v$) (line 2), we first generate the user
representation $\textbf{u}$ (line 7) and item representation $\textbf{v}'$ (line 8), which are
used to compute the click probability $\hat{y}_{uv}$ as
\begin{align}
   \hat{y}_{uv} &= \sigma'(\textbf{u}^{T}\textbf{v}')
\end{align} 
where $\sigma'$ is the sigmoid function.


To optimize \system, we use negative sampling \cite{DBLP:journals/corr/negativesampling} during training.
The objective function is

\vspace{-0.5 pc}
\begin{equation}
    \mathcal{L} = \sum_{u \in \mathcal{U}}(\sum_{v{:}y_{uv=1}}\mathcal{J}(y_{uv},\hat{y}_{uv})-\sum_{i=1}^{N^{u}}\mathbb{E}_{V_{i}  \thicksim P(v_{i})}\mathcal{J}(y_{uv_{i}},\hat{y}_{uv_{i}})) \\ + \lambda \left \| \mathcal{F} \right\|^{2}_{2}
\end{equation}

\noindent
where the first term $\mathcal{J}$ is cross-entropy loss, $P$ is a
negative sampling distribution, and $N^{u}$ is the number of negative samples
for user $u$; $N^{u} = |\{v:y_{uv}=1\}|$, and $P$ follows a uniform
distribution. The second term is the L2 regularizer.

\subsubsection{Fixed-size sampling}

In a real-world knowledge graph, the size of $N(e)$ varies significantly.
In addition, $S^{p}_{u}$ may grow too quickly with the
number of hops. To maintain computational efficiency, we adopt
a fixed-size strategy~\cite{DBLP:journals/corr/abs-1904-12575,DBLP:journals/corr/abs-1806-01973}
and 
sample the set of entities for sections~\ref{User-Oriented Mixing
Information} and \ref{KG-Enhanced User Preference}.  

For Section~\ref{User-Oriented Mixing Information}, we uniformly sample a
fixed-size set of neighbors $\mathcal{N}'(v)$ for each entity $v$, where
$\mathcal{N}'(v) \triangleq \{e|e \thicksim \mathcal{N}(v)\}$ and
$\mathcal{N}(v)$ denotes those entities directly connected to $v$, where
$|\mathcal{N}'(v)|=K_{n}$ and $K_{n}$ is the sampling size of the item neighborhoods and
can be modified.\footnote{We discuss the performance changes when $K_n$
and $K_m$ vary.} 
Also, we do not compute the next-order entity representations
for all entities $e \in \mathcal{G}$, as shown in line 6 of
Algorithm 2, and we sample only a minimal
number of entities to calculate the final entity embedding $\textbf{v}'$.
Per Section~\ref{KG-Enhanced User Preference}, at hop $p$ we sample user
preferences set $\mathcal{S}^{p}_{u}$ to maintain a fixed number of relevant entities,
where $|\mathcal{S}^{p}_{u}|=K_{m}$ and $K_{m}$ is the fixed neighbor sample size, 
which can be modified.$^{5}$

\subsubsection{Stage-wise Training}



To solve the potential issue that the fixed-size sampling strategy may put limitation on the use of all entities, recently stage-wise training has been proposed to collect more entity-relation from KG to approach the panoramic view of the whole neighborhood~\cite{2019arXiv190805611T}. Specifically, in each stage, stage-wise training would resample another set of entities to allow \system to collect more entity information from KG. 
The whole algorithm of stage-wise training is shown in the Algorithm 3 (Line\ref{alg:Learning_Stage-wise Training}).

\newcommand\mycommfont[1]{\footnotesize\ttfamily\textcolor{blue}{#1}}
\SetKwInput{KwInput}{Input}                
\SetKwInput{KwOutput}{Output}              

\begin{algorithm}[h]
\label{alg:Learning_algorithm_KGPI}
\small
\DontPrintSemicolon
\KwInput{Interaction matrix Y, knowledge graph
$\mathcal{G}(\mathcal{E},\mathcal{R})$;}
\KwOutput{Prediction function
$\mathcal{F}(u,v|\Theta, Y, \mathcal{G})$;}
\SetKwFunction{FMain}{ }
\SetKwProg{Fn}{Regular Training}{:}{}
\Fn{\FMain}{
Initialize all parameters; \\
Calculate preference set $\mathcal{S}_{u}$ for each user $u$; \\
Map neighborhood sample $\mathcal{N}'(v)$ for each node; \\  
 \While{\system has not converged}{
    \For{$(u,v)$ in $\mathcal{Y}$}{
        \textbf{u} $\gets$ KGUR($u$);\\
        \textbf{v}$'$ $\gets$ MixLayer($v,\textbf{u}$);\\
        Calculate predicted probability $ \hat{y}_{uv}$ = $f$(\textbf{u},\textbf{v}$'$); \\
        Update parameters by gradient descent;\\
    }
}
}

\SetKwFunction{FMain}{ }
\SetKwProg{Fn}{Stage-wise Training}{:}{}
\Fn{\FMain}{
    \label{alg:Learning_Stage-wise Training}
    Initialize all parameters;\\
    Save embedding of $\mathcal{G}(\mathcal{E},\mathcal{R})$;\\
    \While{\system has not converged}{
        Initialize all parameters;\\
        Load previous embedding of  $\mathcal{G}(\mathcal{E},\mathcal{R})$;\\
    Re-sample $\mathcal{S}_{u}$ and $\mathcal{N}'(v)$ according to Eq.~(3)--(4); \\
    Calculate Eq.~(5)--(10); \\
    Save best embedding of $\mathcal{G}(\mathcal{E},\mathcal{R})$;\\
    }
}
\caption{\system Learning}
\end{algorithm}

\subsubsection{Time Complexity Analysis} Per batch, the time cost for \system
mainly comes from generating KG-enhanced user representation and the mixing
layer. The user representation generation has a computational complexity of
$O(l_{\mathrm{p}}K_{m}s^{2})$ to calculate the relevance probability $k_{i}$
for total of $l_{\mathrm{p}}$ layers. The mixing layer has a computational
complexity of $O({K_{n}}^{l_{\mathrm{w}}l_{\mathrm{d}}}s^{2})$ to aggregate
through the deep layer $l_{\mathrm{d}}$ and wide layer $l_{\mathrm{w}}$. 
The overall training complexity of \system is thus $O(l_{\mathrm{p}}K_{m}s^{2} +
{K_{n}}^{l_{\mathrm{w}}l_{\mathrm{d}}}s^{2})$.


Compared with other GNN-based recommendation models such as RippleNet, KGCN,
and KGAT, \system achieves a comparable computation complexity level. Below,
we set their layers to $l$ and the sampling number to $K$
for simplicity. The computational complexities of RippleNet and KGCN are
$O(lKs^{2})$ and $O(K^{l}s^{2})$ respectively. This is at the same level as ours
because ${l_{w}l_{d}}$ is a special case of $l$.
However, for KGAT, without the sampling strategy, its attention embedding
propagation part should globally update the all entities in graph, and its
computational complexity is $O(l|\mathcal{G}|s^2)$.

We conducted experiments to compare the training speed of the proposed
\system and others on an RTX-2080 GPU. Empirically, \system, RippleNet, KGCN,
and KGAT take around 6.5s, 5.8s, 3.7s, and 550s respectively to iterate all
training user-item pairs in the Amazon-Book dataset. We see that \system has a
time consumption comparable with RippleNet and KGCN, but KGAT is inefficient
because of the whole-graph updates.

\section{Experiments and Results}
\label{sec:exp}

In this section, we introduce the datasets, baseline models, and experiment
setup, followed by the results and discussion.


\begin{table}[t]
\small
\centering
\begin{tabular}{c|cccccc}  & 
\multicolumn{1}{c}{ML-1M} & 
\multicolumn{1}{c}{LFM-1b}  & 
\multicolumn{1}{c}{AZ-book}\\
\hline 
Users & 6,036  & 12,134 & 6,969  \\ 
Items & 2,445 & 15,471 & 9,854 \\ 
Interactions & 753,772  & 2,979,267 & 552,706 \\ 
Avg user clicks & 124.9 & 152.3 & 79.3 \\ 
Avg clicked items  & 308.3  & 119.4 & 56.1 \\ \hline 
KG source & Microsoft Satori & Freebase &Freebase \\ 
KG entities & 182,011  & 106,389 & 113,487  \\ 
KG relations & 12 & 9 & 39 \\ 
KG triples & 1,241,995 & 464,567 & 2,557,746  \\
\end{tabular}
\caption{Dataset statistics}
\label{tb:datasetstatistic}
\vspace{-2.0 pc}
\end{table}

\subsection{Datasets}

In the evaluation, we utilized three real-world datasets:
ML-1M, LFM-1b, and AZ-book which are publicly
available~\cite{DBLP:journals/corr/abs-1803-03467,DBLP:journals/corr/abs-1904-12575,DBLP:journals/corr/abs-1905-07854}.
We compared \system with models working on these datasets coupled with various
KGs, which were built in different ways. For
ML-1M, its KGs were built by Microsoft Satori where the confidence level was set
to greater than 0.9. The KGs of LFM-1b and AZ-book were built by title matching
as described in
\cite{DBLP:journals/corr/abs-1807-11141}. The
statistics of the three datasets are shown in Table~\ref{tb:datasetstatistic},
and their descriptions are as follows:

\begin{itemize}
\item \textbf{MovieLens-1M} A benchmark dataset for movie recommendations
with approximately 1 million explicit
ratings (ranging from 1 to 5) on a total of 2,445 items from 6,036 users. 
\item \textbf{LFM-1b 2015} A music dataset which records artists, albums,
tracks, and users, as well as individual listening events and contains about 3
million explicit rating records on 15,471 items from
12,134 users.
\item \textbf{Amazon-book} Records user preferences on book products. It
records information about users, items, ratings, and event timestamps. This
dataset contains about half a million explicit rating records on
a total of 9,854 items from 7,000 users.
\end{itemize}

We transformed the ratings into binary feedback, where
each entry was marked as 1 if the item had been rated by users;
otherwise, it was marked as 0. The rating threshold of ML-1M was 4;
that is, if the item was rated less than 4 by the user, the entry was set to 0.
For LFM-1b and AZ-book, the entry was marked as 1 if user-item interaction was
observed. 
To ensure dataset quality, we applied a $g$-core setting, i.e., we retained users
and items with at least $g$ interactions. For AZ-book and LFM-1b, $g$ was set to
20.

\subsection{Baseline Models}

To evaluate the performance, we 
compared the proposed \system with the following baselines, CF-based (FM and
NFM), regularization-based (CKE), path-based (MCRec), and graph neural
network-based (GC-MC, KGCN, RippleNet, and KGAT) methods.

\begin{itemize}


\item \textbf{FM}~\cite{Rendle:2011:FCR:2009916.2010002} A widely used
factorization approach for modeling feature interaction. In our evaluations,
we concatenated IDs of user, item, and related KG knowledge
as input features.

\item \textbf{NFM}~\cite{10.1145/3077136.3080777} A
factorization-based method which seamlessly combines the linearity and
non-linearity of neural networks in modeling user-item interaction. Here, to
enrich the representation of an item, we followed~\cite{10.1145/3077136.3080777} and fed NFM with the embeddings of its connected entities on KG.

\item \textbf{GC-MC}~\cite{vdberg2017graph}
A graph-based auto-encoder framework for matrix completion. GC-MC is a
GCN-based recommendation model which encodes a user-item bipartite graph by graph
convolutional matrix completion. We used implicit user-item interaction to
create a user-item bipartite graph.


\item \textbf{CKE}~\cite{Zhang:2016:CKB:2939672.2939673} A regularization-based
method. CKE combines structural, textual, and visual knowledge and learns
jointly for recommendation. We used structural knowledge and recommendation
component as input.

\item \textbf{MCRec}~\cite{Hu:2018:LMB:3219819.3219965_path} A co-attentive
model which requires finer meta paths, which connect users and items, 
to learn context representation. The co-attention
mechanism improves the representations for meta-path-based context, users, and
items in a mutually enhancing way.

\item \textbf{KGCN}~\cite{DBLP:journals/corr/abs-1904-12575} Utilizes GCN to
collect high-order neighborhood information from the KG. To find the neighborhood
which the user may be more interested in, it uses the user representation to
attend on different relations to calculate the weight of the neighborhood.

\item \textbf{RippleNet}~\cite{DBLP:journals/corr/abs-1803-03467} A
memory-network-like approach which represents the user by his or her
related items. RippleNet uses all relevant entities in the KG to propagate the user's 
representation for recommendation.

\item \textbf{KGAT}~\cite{DBLP:journals/corr/abs-1905-07854} A GNN-based
recommendation model equipped with a graph attention network. It
uses a hybrid structure of the knowledge graph and user-item graph as a collaborative
knowledge graph. KGAT employs an attention mechanism to discriminate the
importance of neighbors and outperforms several state-of-the-art methods.

\end{itemize}

\vspace{-0.3 pc}

\begin{table*}[t]
\small
\centering
\caption{$\mathit{AUC}$ and $\mathit{ACC}$ results in CTR prediction on all datasets.}
\vspace{-0.5 pc}
\scalebox{1.0}{
\begin{tabular}{ccccccc}
\hline \multicolumn{1}{c}{\multirow{2}{*}{Model}} &
\multicolumn{2}{c}{ML-1M}&
\multicolumn{2}{c}{LFM-1b}&
\multicolumn{2}{c}{AZ-book} \\
\cmidrule(lr){2-3} \cmidrule(lr){4-5} \cmidrule(lr){6-7}  &
\multicolumn{1}{c}{\emph{AUC}} & \multicolumn{1}{c}{\emph{ACC}} &
\multicolumn{1}{c}{\emph{AUC}} & \multicolumn{1}{c}{\emph{ACC}} &
\multicolumn{1}{c}{\emph{AUC}} & \multicolumn{1}{c}{\emph{ACC}} \\
\hline \multicolumn{1}{c}{{FM}} & \multicolumn{1}{c}{{.9101 (-2.3\%)}} & \multicolumn{1}{c}{{.8328 (-2.9\%)}} & \multicolumn{1}{c}{{.9052 (-6.3\%)}} & \multicolumn{1}{c}{{.8602 (-5.6\%)}} & \multicolumn{1}{c}{{.7860 (-10.2\%)}} & \multicolumn{1}{c}{{.7107 (-10.4\%)}} \\
\multicolumn{1}{c}{{NFM}} & \multicolumn{1}{c}{{.9167 (-1.6\%)}} &
\multicolumn{1}{c}{{.8420 (-1.8\%)}} & \multicolumn{1}{c}{{.9301 (-3.7\%)}} & \multicolumn{1}{c}{{.8825 (-3.2\%)}} & \multicolumn{1}{c}{{.8206 (-6.2\%)}} & \multicolumn{1}{c}{{.7474 (-5.8\%)}}  \\
\multicolumn{1}{c}{{CKE}} & \multicolumn{1}{c}{{.9095 (-2.4\%)}} &
\multicolumn{1}{c}{{.8376 (-2.3\%)}} & \multicolumn{1}{c}{{.9035 (-6.5\%)}} & \multicolumn{1}{c}{{.8591 (-5.7\%)}} & \multicolumn{1}{c}{{.8070 (-7.8\%)}} & \multicolumn{1}{c}{{.7227 (-8.9\%)}}  \\
\multicolumn{1}{c}{{MCRec}} & \multicolumn{1}{c}{{.8970 (-3.7\%)}} &
\multicolumn{1}{c}{{.8262 (-3.6\%)}} & \multicolumn{1}{c}{{.8920 (-7.6\%)}} & \multicolumn{1}{c}{{.8428 (-7.5\%)}} & \multicolumn{1}{c}{{.7925 (-9.4\%)}} & \multicolumn{1}{c}{{.7217 (-9.1\%)}}  \\
\multicolumn{1}{c}{{KGNN}} & \multicolumn{1}{c}{{.9093 (-2.4\%)}} &
\multicolumn{1}{c}{{.8338 (-2.7\%)}} & \multicolumn{1}{c}{{.9171 (-5.0\%)}} & \multicolumn{1}{c}{{.8664 (-4.9\%)}} & \multicolumn{1}{c}{{.8043 (-8.1\%)}} & \multicolumn{1}{c}{{.7291 (-8.1\%)}} \\
\multicolumn{1}{c}{{RippleNet}} & \multicolumn{1}{c}{{.9208 (-1.2\%)}} &
\multicolumn{1}{c}{{.8435 (-1.6\%)}} & \multicolumn{1}{c}{{.9421 (-2.5\%)}} & \multicolumn{1}{c}{{.8887 (-2.5\%)}} & \multicolumn{1}{c}{{.8234 (-5.9\%)}} & \multicolumn{1}{c}{{.7486 (-5.7\%)}} \\
\multicolumn{1}{c}{{KGAT}} & \multicolumn{1}{c}{{.9222 (-1.2\%)}} &
\multicolumn{1}{c}{{.8489 (-1.0\%)}} & \multicolumn{1}{c}{{.9384 (-2.8\%)}} & \multicolumn{1}{c}{{.8771 (-3.7\%)}} & \multicolumn{1}{c}{{.8555 (-2.2\%)}} & \multicolumn{1}{c}{{.7793 (-1.8\%)}} \\
\multicolumn{1}{c}{{GC-MC}} & \multicolumn{1}{c}{{.9005 (-3.4\%)}} &
\multicolumn{1}{c}{{.8197 (-4.4\%)}} & \multicolumn{1}{c}{{.9204 (-4.7\%)}} & \multicolumn{1}{c}{{.8723 (-4.3\%)}} & \multicolumn{1}{c}{{.8177 (-6.5\%)}} & \multicolumn{1}{c}{{.7347 (-7.4\%)}} \\
\hline
\multicolumn{1}{c}{{\system}} & \multicolumn{1}{c}{{\textbf{.9318}* (\%)}} &
\multicolumn{1}{c}{{\textbf{.8573}* (\%)}} & \multicolumn{1}{c}{{\textbf{.9658}* (\%)}} & \multicolumn{1}{c}{{\textbf{.9112}* (\%)}} & \multicolumn{1}{c}{{\textbf{.8749}* (\%)}} & \multicolumn{1}{c}{{\textbf{.7935}* (\%)}} \\
\hline \multicolumn{6}{@{} l}
{\scriptsize Note: * indicates statistically significant improvements over the
best baseline by an unpaired two-sample $t$-test with $p$-value = 0.01.}
\end{tabular}}
\label{tb:KGPH_CTR_result}
\vspace{-0.5 pc}
\end{table*}

\begin{figure*}[htb]
\centering
\includegraphics[scale=0.38, trim={0 0 0 0}]{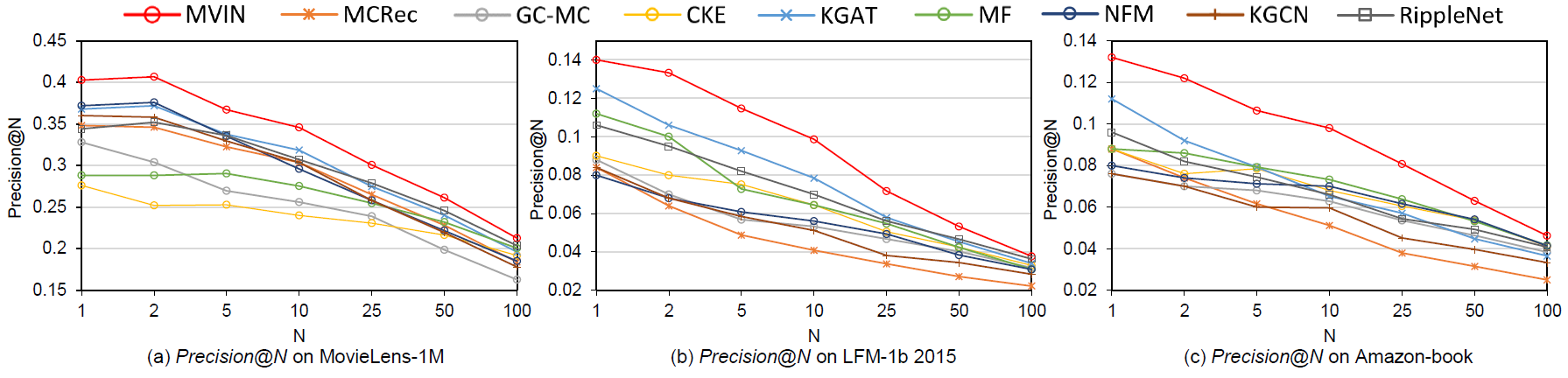}
\vspace{-1.5 pc}
\caption{$Precision@N$ results in top-$N$ recommendation.}
\label{fig:KGPH_top_K}
\vspace{-0.8 pc}
\end{figure*}

\subsection{Experiments} 

\subsubsection{Experimental Setup}  
~\label{Experimental_Setup}
For \system, $l_\mathrm{p}$ = 2, $l_\mathrm{w}$ = 1, $l_\mathrm{d}$ = 2, $K_m$
= 64, $K_n$ = 8, $\lambda$ = $1\times {10}^{-7}$ for ML-1M; $l_\mathrm{p}$ = 1,
$l_\mathrm{w}$ = 1, $l_\mathrm{d}$ = 2, $K_m$ = 64, $K_n$ = 4, $\lambda$ =
$5\times {10}^{-8}$ for LFM-1b; $l_\mathrm{p}$ = 2, $l_\mathrm{w}$ = 2,
$l_\mathrm{d}$ = 2, $K_m$ = 16, $K_n$ = 8, $\lambda$ = $1\times {10}^{-7}$ for
AZ-book; We set function $\sigma$ as ReLU. The embedding size was fixed to 16
for all models except 32 for KGAT because it stacks propagation layers for
final output. For stage-wise training, average early stopping stage number is 7, 7, 5 for ML-1M, LFM-1b and AZ-book, respectively. For all models, the hyperparameters were determined by optimizing $\mathit{AUC}$ on
a validation set. For all models, the learning rate $\eta$ and regularization
weight were selected from [$2\times {10}^{-2}$, $1\times {10}^{-2}$, $5\times
{10}^{-3}$, $5\times {10}^{-4}$, $2\times {10}^{-4}$] and from [$1\times
{10}^{-4}$, $1\times{10}^{-5}$, $2\times{10}^{-5}$, $2\times {10}^{-7}$,
$1\times {10}^{-7}$, $5\times{10}^{-8}$], respectively. For MCRec, to define
several types of meta paths, we manually selected user-item-attribute item
meta paths for each dataset and set the hidden layers as in~\cite{Hu:2018:LMB:3219819.3219965_path}.
For KGAT, we set the depth to 2 and layer size to [16,16]. For RippleNet, we
set the number of hops to 2 and the sampling size to 64 for each dataset. For
KGCN, we set the number of hops to 2, 2, 1 and sampling size to 4, 8, 8 for
ML-1M, AZ-book, and LFM-1b, respectively. Other hyperparameters were optimized according to validation result.



\subsubsection{Experimental Results}  

Table~\ref{tb:KGPH_CTR_result} and Figure~\ref{fig:KGPH_top_K} are the results
of \system and the baselines, respectively (FM, NFM, CKF, GC-MC, MCRec, RippleNet,
KGCN, KGAT), in click-through rate (CTR) prediction, i.e., taking a user-item pair as input and predicting the probability of the user engaging with the item. We adopt $\mathit{AUC}$ and $\mathit{ACC}$, which are widely used in binary classification problems, to evaluate the performance of CTR prediction. For those of top-$N$ recommendation, selecting $N$ items with highest predicted click probability for each user and choose $\mathit{Precision@N}$ to evaluate the recommended sets. We have the following observations:

\vspace{-0.3 pc}


\begin{itemize}

\item \system yields the best performance of all the datasets
and achieves $\mathit{AUC}$ performance gains of 1.2\% , 2.5\%, and 2.2\% on ML-1M,
LFM-1b, and AZ-book, respectively. Also, \system achieves outstanding
performance in top-$N$
recommendation, as shown in Figure~\ref{fig:KGPH_top_K}. 



\item The two path-based baselines RippleNet and KGAT
outperform the two CF-based methods FM and NFM,
indicating that KG is helpful for recommendation. Furthermore, although
RippleNet and KGAT achieve excellent performance, they still do not
outperform \system. This is because RippleNet neither incorporates user click
history items $h^{1}_{i}$ into user representation nor does it introduces high-order
connectivities, and KGAT does not mix GCN layer information and not  v consider user
preferences when collecting KG information.

\item For the other baselines KGCN and MCRec, their relatively bad performance is
attributed to their not fully utilizing information from user click items. In contrast, MVIN would first enrich a user representation by user  click items and all relevant entities in KG and then weighted the nearby entities and emphasize the most important ones. Also, KGCN only uses GCN in each layer, which does not allow 
contrast on neighborhood layers. Furthermore, MCRec requires finer defined meta paths,
which requires manual labor and domain knowledge.


\item To our surprise, the CF-based NFM achieves good performance on
LFM-1b and AZ-book, even outperforming the KG-aware baseline KGCN, and achieves
results comparable to RippleNet. Upon investigation, we found that this is because we
enriched its item representation by feeding the embeddings of its connected
entities. In addition, NFM's design involves modeling higher-order and
non-linear feature interactions and thus better captures
interactions between user and item embeddings. These observation conform 
to~\cite{DBLP:journals/corr/abs-1905-07854}.

\item The regularization-based CKE is outperformed by NFM.
CKE does not make full use of the KG because it is only regularized by correct
triplets from KG. Also, CKE neglects high-order connectivities.


\item Although GC-MC has introduced high-order connectivity into user and
item representations, it achieves comparably weak performance as it only
utilizes a user-item bipartite graph and ignores semantic information between KG
entities. 

 
\vspace{-0.8 pc}
 
\end{itemize}

\subsection{Study of \system}

\begin{table*}[t]
\small
\centering
\caption{\system ablation study results. We evaluate using $\mathit{AUC}$ in
CTR prediction on all datasets and show the effect of the proposed methods. 
User-oriented modules contain entity projection (UO(e)), relation attention
(UO(r)), and KG-enhanced user-oriented information (UO(k)). Mixing
layer has deep (ML(d)) and wide (ML(w)) parts. Stage-wise training (SW)
is used as well. * indicates statistically significant improvements by an unpaired
two-sample $t$-test with $p$-value = 0.01.}

\vspace{-0.2 pc}
\scalebox{1.0}{
\begin{tabular}{cp{1cm}p{3cm}p{3cm}p{3cm}p{3cm}p{3cm}p{3cm}p{3cm}ccccccc}
\hline
\multicolumn{1}{c}{\multirow{2}{*}{Ablation component}}&
\multicolumn{6}{c}{Components}&
\multicolumn{1}{c}{ML-1M}&
\multicolumn{1}{c}{LFM-1b}&
\multicolumn{1}{c}{AZ-book} \\

\cmidrule(lr){2-7}
\cmidrule(lr){8-10} &
\multicolumn{1}{c}{UO(e)}&
\multicolumn{1}{c}{UO(r)}&
\multicolumn{1}{c}{UO(k)}&
\multicolumn{1}{c}{ML(w)}& 
\multicolumn{1}{c}{ML(d)}&
\multicolumn{1}{c}{ SW }&
\multicolumn{1}{c}{\emph{AUC}} &
\multicolumn{1}{c}{\emph{AUC}} &
\multicolumn{1}{c}{\emph{AUC}} \\ 
\hline 
N/A
&
\multicolumn{1}{c}{{{\color{blue} \CheckmarkBold}}} & \multicolumn{1}{c}{{{\color{blue} \CheckmarkBold}}} & \multicolumn{1}{c}{{{\color{blue} \CheckmarkBold}}} & 
\multicolumn{1}{c}{{{\color{blue} \CheckmarkBold}}} & 
\multicolumn{1}{c}{{{\color{blue} \CheckmarkBold}}} & 
\multicolumn{1}{c}{{{\color{blue} \CheckmarkBold}}} &  \multicolumn{1}{c}{{\textbf{.9318}}} & \multicolumn{1}{c}{{\textbf{.9658}}} & \multicolumn{1}{c}{{\textbf{.8739}}}\\
w/o UO(e)
&
\multicolumn{1}{c}{{}} & 
\multicolumn{1}{c}{{{\color{blue} \CheckmarkBold}}} & 
\multicolumn{1}{c}{{{\color{blue} \CheckmarkBold}}} & 
\multicolumn{1}{c}{{{\color{blue} \CheckmarkBold}}} & 
\multicolumn{1}{c}{{{\color{blue} \CheckmarkBold}}} & 
\multicolumn{1}{c}{{{\color{blue} \CheckmarkBold}}} & \multicolumn{1}{c}{{.9299 (-0.2\%)}} & \multicolumn{1}{c}{{.9617 (-0.4\%)*}} & \multicolumn{1}{c}{{.8672 (-0.8\%)*}} \\ 
w/o UO(r)
&
\multicolumn{1}{c}{{{\color{blue} \CheckmarkBold}}} & 
\multicolumn{1}{c}{{}} & 
\multicolumn{1}{c}{{{\color{blue} \CheckmarkBold}}} & 
\multicolumn{1}{c}{{{\color{blue} \CheckmarkBold}}} & 
\multicolumn{1}{c}{{{\color{blue} \CheckmarkBold}}} &
\multicolumn{1}{c}{{{\color{blue} \CheckmarkBold}}} & \multicolumn{1}{c}{{.9305 (-0.1\%)}} & \multicolumn{1}{c}{{.9638 (-0.2\%)}} & \multicolumn{1}{c}{{.8705 (-0.4\%)*}} \\
w/o UO(k)
&
\multicolumn{1}{c}{{{\color{blue} \CheckmarkBold}}} & 
\multicolumn{1}{c}{{{\color{blue} \CheckmarkBold}}} & 
\multicolumn{1}{c}{{{}}} & 
\multicolumn{1}{c}{{{\color{blue} \CheckmarkBold}}} & 
\multicolumn{1}{c}{{{\color{blue} \CheckmarkBold}}} & 
\multicolumn{1}{c}{{{\color{blue} \CheckmarkBold}}} & \multicolumn{1}{c}{{.9247 (-0.7\%)*}} & \multicolumn{1}{c}{{.9598 (-0.6\%)*}} & \multicolumn{1}{c}{{.8573 (-1.8\%)*}} \\
w/o ML(w)
&
\multicolumn{1}{c}{{{\color{blue} \CheckmarkBold}}} & 
\multicolumn{1}{c}{{{\color{blue} \CheckmarkBold}}} & 
\multicolumn{1}{c}{{{\color{blue} \CheckmarkBold}}} & 
\multicolumn{1}{c}{{}} & 
\multicolumn{1}{c}{{{\color{blue} \CheckmarkBold}}} & 
\multicolumn{1}{c}{{{\color{blue} \CheckmarkBold}}} & \multicolumn{1}{c}{{.9289 (-0.3\%)*}} & \multicolumn{1}{c}{{.9621 (-0.4\%)*}} & \multicolumn{1}{c}{{.8683 (-0.6\%)*}} \\
w/o ML
&
\multicolumn{1}{c}{{}} & 
\multicolumn{1}{c}{{}} & 
\multicolumn{1}{c}{{}} & 
\multicolumn{1}{c}{{}} & 
\multicolumn{1}{c}{{{}}} & 
\multicolumn{1}{c}{{{\color{blue} \CheckmarkBold}}} & \multicolumn{1}{c}{{.9283(-0.4\%)*}} & \multicolumn{1}{c}{{.9613 (-0.5\%)*}} & \multicolumn{1}{c}{{.8637 (-1.2\%)*}} \\
w/o SW
&
\multicolumn{1}{c}{{{\color{blue} \CheckmarkBold}}} & 
\multicolumn{1}{c}{{{\color{blue} \CheckmarkBold}}} & 
\multicolumn{1}{c}{{{\color{blue} \CheckmarkBold}}} & 
\multicolumn{1}{c}{{{\color{blue} \CheckmarkBold}}} & 
\multicolumn{1}{c}{{{\color{blue} \CheckmarkBold}}} & 
\multicolumn{1}{c}{{}} &  \multicolumn{1}{c}{{.9276 (-0.5\%)*}} & \multicolumn{1}{c}{{.9567 (-0.9\%)*}} & \multicolumn{1}{c}{{.8642 (-1.1\%)*}} \\
\hline
\end{tabular}}
\label{tb:KGPI_abla_result}
\vspace{-0.8 pc}
\end{table*}

We conducted an ablation study to verify the effectiveness of the proposed components.
We also provide an in-depth exploration of the entity view. 

\subsubsection{User-Oriented Information}
The ablation study results are shown in Table~\ref{tb:KGPI_abla_result}.
After removing the proposed user-oriented relation attention UO(r) and
user-oriented entity projection UO(e) modules, 
MVIN$_{\mathrm{w/o\,UO(r)}}$ and MVIN$_{\mathrm{w/o\,UO(e)}}$ perform worse
than \system in all datasets. Thus
considering user preferences when aggregating entities and relations in
KG improves recommendation results.




\begin{figure}[htb]
\centering
\includegraphics[scale=0.285, trim={20 0 0 0}]{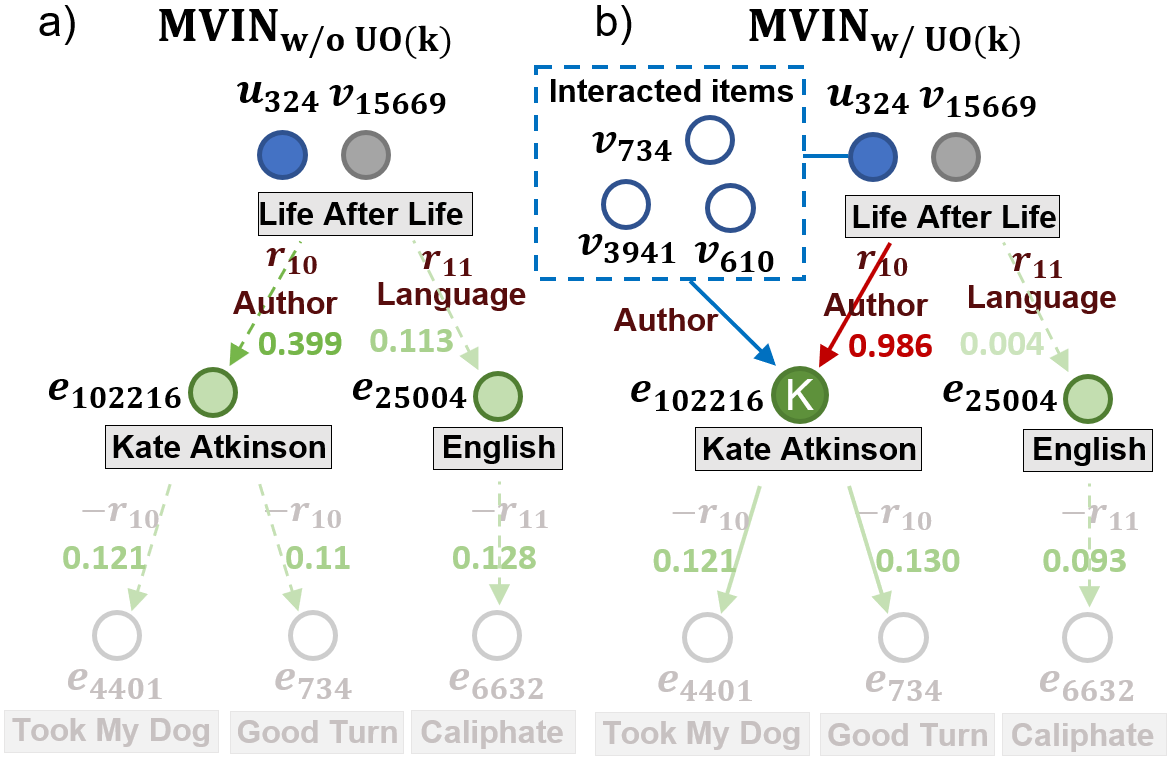}

\vspace{-0.5 pc}
\caption{Attention visualization. We compare the attention weights between
a) MVIN$_{\mathrm{w/o\,UO(k)}}$ and (b) \system. Results show that when 
information on user-interacted items is provided, (b) \system pays more
attention to \textit{Kate Atkinson}, the author which the user may be 
interested in.}
\label{fig:MVIN_uo_k_case_st}

\vspace{-0.5 pc}
\end{figure}


\subsubsection{KG-enhanced User-Oriented Information}
\label{Effect of KG-enhanced User-Orient Information}
To enhance user-oriented information,
we enrich the user representation using KG information as a pre-processing step. Here,
we denote \system without KG-enhanced user-oriented information as
MVIN$_{\mathrm{w/o\,UO(k)}}$.
We compare the performance of \system and MVIN$_{\mathrm{w/o\,UO(k)}}$.
Table~\ref{tb:KGPI_abla_result} shows that the former outperforms the latter by
a large margin, which confirms that KG-enhanced user representation improves
user-oriented information.

Moreover, we conducted a case study to understand the effect of KG-enhanced
user-oriented information incorporated with user-entity interaction. 
Given the attention weights learned by MVIN $_{\mathrm{w/o\,UO(k)}}$ in
Figure~\ref{fig:MVIN_uo_k_case_st}(a), user $\mathrm{u}_{324}$ puts only slightly more value
on the author of \emph{Life After Life}. However,
Figure~\ref{fig:MVIN_uo_k_case_st}(b) shows that \system puts much more attention
on the author when information on user-interacted items is provided.
Furthermore, in Figure~\ref{fig:MVIN_uo_k_case_st}(b), we find user
$\mathrm{u}_{324}$'s interacted items---$v_{734}$ (\textit{One Good Turn}),
$v_{3941}$ (\textit{Behind the Scenes at the Museum}), and $v_{610}$
(\textit{Case Histories})---are all written by \textit{Kate
Atkinson}. This demonstrates that \system outperforms
MVIN$_{\mathrm{w/o\,UO(k)}}$ because it captures the most important view that
user $\mathrm{u}_{324}$ sees: item \emph{Life After Life}, a book by \emph{Kate
Atkinson}.





  

\subsubsection{Mixing layer-wise GCN information} In the mixing layer, 
the wide part ML(w) allows \system to represent general layer-wise
neighborhood mixing. To study the effect of ML(w), we remove the wide part
from \system, denoted as MVIN$_{\mathrm{w/o\,ML(w)}}$. 
Table~\ref{tb:KGPI_abla_result} shows a drop in performance, suggesting
that the mixing of features from different distances improves
recommendation performance.

\begin{figure}[htb]
\vspace{-0.8 pc}
\centering
\includegraphics[scale=0.29, trim={0 0 0 0}]{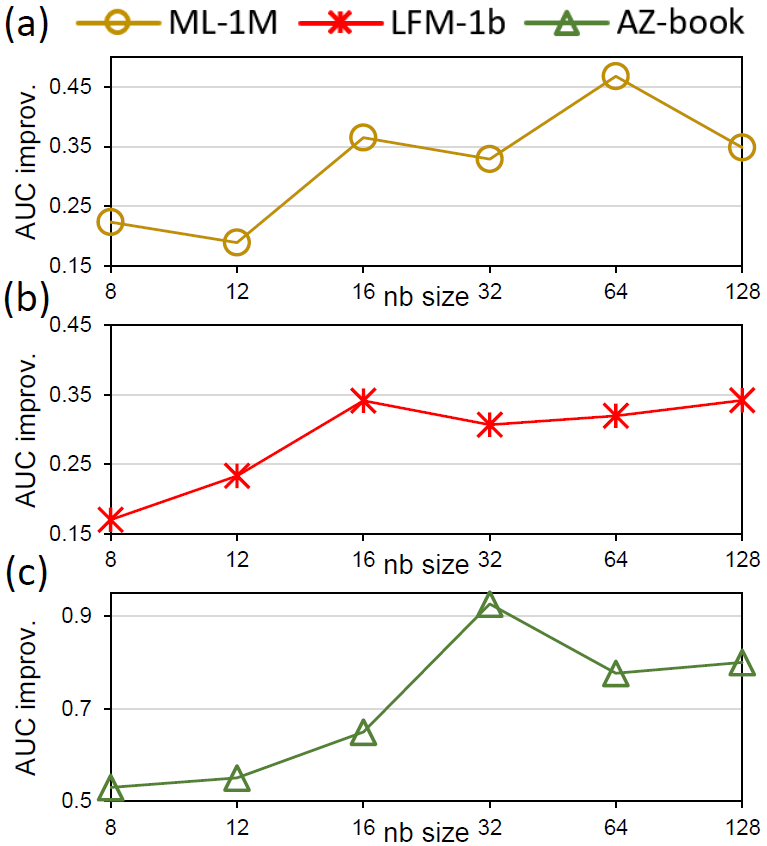}
\vspace{-0.5 pc}
\caption{$\mathit{AUC}$ improvement from MVIN$_{\mathrm{w/o\,ML(w)}}$ to MVIN for
different neighborhood sampling sizes $K_n$, where the preference set size $K_m$
is set to 16. With a large $K_n$, the performance gap between \system and
MVIN$_{\mathrm{w/o\,ML(w)}}$ increases, indicating the indispensability of ML(w) for
large $K_n$.}
\vspace{-0.8 pc}
\label{fig:KGPI_effect_of_wl}
\end{figure}

\subsubsection{Mixing layer-wise GCN information (at high neighbor
sampling size $K_n$)}
\label{Effect_of_mixing_layer-wise_GCN_information_high_sampling_size}
It has shown that in homogeneous graphs the benefit of the mixing
layer depends on the homophily level.\footnote{The homophily level indicates the
likelihood of a node forming a connection to a neighbor with the same label.} In
\system, the mixing layer works in KGs, i.e., heterogeneous graphs; we also
investigate its effect (ML(w)) at different sampling sizes of neighborhood
nodes $K_n$. With a large $K_n$, entities in KG connect to more different
entities, which is similar to a low homophily level.
Figure~\ref{fig:KGPI_effect_of_wl} shows that ML(w) is effective in
heterogeneous graphs. In addition, $K_n$ increases the performance gap between
MVIN and MVIN$_{\mathrm{w/o\,ML(w)}}$. We conclude that the mixing
layer not only improves MVIN performance but is indispensable for
large $K_n$. 






\subsubsection{High-order connectivity information}
In addition to the wide part in the mixing layer, the proposed deep part 
allows \system to aggregate high-order connectivity information. 
Figure~\ref{tb:KGPI_abla_result} shows that after removing the mixing layer (ML),
MVIN$_{\mathrm{w/o\,ML}}$ performs poorly compared to \system, demonstrating
the significance of high-order connectivity. This observation is consistent
with~\cite{DBLP:journals/corr/abs-1905-07854,DBLP:journals/corr/abs-1905-04413,DBLP:journals/corr/abs-1806-01973}.


\subsubsection{Stage-wise Training}
Removing the stage-wise training (SW) shown by MVIN$_{\mathrm{w/o\,SW}}$
deteriorates performance, showing that stage-wise training helps \system
achieve better performance by collecting more entity relations from the KG to
approximate a panoramic view of the whole neighborhood. Note
that compared to KGAT, the state-of-the-art baseline model which samples the whole neighbor
entities in KG, MVIN$_{\mathrm{w/o\,SW}}$ refers to a limited number of
entities in KG but still significantly outperforms all
baselines (at $p$-value = 0.01), which confirms
again the effectiveness of the proposed \system. 







\begin{table}[t]
\small
\centering
\caption{$\mathit{AUC}$ of \system with different preference set size $K_m$ and neighbor
sampling size $K_n$.}
\vspace{-0.5 pc}
\scalebox{1.0}{
\begin{tabular}{p{2.5cm}|ccccc}
\hline \multicolumn{1}{c|}{\multirow{1}{*}{$K_m$ size
($K_n=4$)}} & \multicolumn{1}{c}{{4}} &
\multicolumn{1}{c}{{8}} & \multicolumn{1}{c}{{16}}&
\multicolumn{1}{c}{{32}} & \multicolumn{1}{c}{{64}} \\ \hline \multicolumn{1}{c|}{ML-1M}& \multicolumn{1}{c}{{.9210}} &
\multicolumn{1}{c}{{.9247}} & \multicolumn{1}{c}{{.9255}} &
\multicolumn{1}{c}{{.9269}} & \multicolumn{1}{c}{{\textbf{.9276}}}  \\ 
\multicolumn{1}{c|}{LFM-1b}  & \multicolumn{1}{c}{{.9299}} &
\multicolumn{1}{c}{{.9368}} & \multicolumn{1}{c}{{.9433}} &
\multicolumn{1}{c}{{.9498}} & \multicolumn{1}{c}{{\textbf{.9567}}} \\
\multicolumn{1}{c|}{AZ-book} 
& \multicolumn{1}{c}{{.8508}}
& \multicolumn{1}{c}{{.8613}}
& \multicolumn{1}{c}{{.8616}} 
& \multicolumn{1}{c}{{\textbf{.8642}}}
& \multicolumn{1}{c}{{.8631}}  \\
\hline \multicolumn{1}{c|}{\multirow{1}{*}{$K_n$ size ($K_m=16$)}} & \multicolumn{1}{c}{{4}} &
\multicolumn{1}{c}{{8}} & \multicolumn{1}{c}{{16}}&
\multicolumn{1}{c}{{32}} & \multicolumn{1}{c}{{64}} \\ \hline \multicolumn{1}{c|}{ML-1M} 
& \multicolumn{1}{c}{{.9246}} 
& \multicolumn{1}{c}{{.9254}}
& \multicolumn{1}{c}{{.9258}}
& \multicolumn{1}{c}{{\textbf{.9264}}}
& \multicolumn{1}{c}{{.9252}}\\ 
\multicolumn{1}{c|}{LFM-1b}  
& \multicolumn{1}{c}{{.9427}} 
& \multicolumn{1}{c}{{.9430}}
& \multicolumn{1}{c}{{\textbf{.9433}}}
& \multicolumn{1}{c}{{.9429}}
& \multicolumn{1}{c}{{.9415}} \\
\multicolumn{1}{c|}{AZ-book}
& \multicolumn{1}{c}{{.8590}} 
& \multicolumn{1}{c}{{.8601}} 
& \multicolumn{1}{c}{{.8594}} 
& \multicolumn{1}{c}{{\textbf{.8610}}}
& \multicolumn{1}{c}{{.8593}}\\
\hline
\end{tabular}}
\label{tb:KGPH_by_me_preference}
\end{table}


\begin{table}[t]
\small
\centering
\caption{$\mathit{AUC}$ of \system with different number of $l_\mathrm{p}$,
$l_\mathrm{w}$, and $l_\mathrm{d}$ hops, where $K_m$ is set to 16. For the
Propagation layer, 0 hops denotes that only the user-clicked items $h^{1}_{i}$ are
utilized.} 
\begin{tabular}{p{2.5cm}|cccc}
\hline \multicolumn{1}{c|}{\multirow{1}{*}{$l_\mathrm{p}$ hops}} & \multicolumn{1}{c}{{0}} & \multicolumn{1}{c}{{1}} & \multicolumn{1}{c}{{2}} &
\multicolumn{1}{c}{{3}} \\ \hline \multicolumn{1}{c|}{ML-1M} &
\multicolumn{1}{c}{{.9257}} &
\multicolumn{1}{c}{{\textbf{.9262}}} & \multicolumn{1}{c}{{.9233}} & \multicolumn{1}{c}{{.9244}} \\ 
\multicolumn{1}{c|}{LFM-1b}  &
\multicolumn{1}{c}{{.9317}} & 
\multicolumn{1}{c}{{\textbf{.9438}}} & \multicolumn{1}{c}{{.9429}} & \multicolumn{1}{c}{{.9415}} \\
\multicolumn{1}{c|}{AZ-book} &
\multicolumn{1}{c}{{.8557}}  & 
\multicolumn{1}{c}{{\textbf{.8576}}} & \multicolumn{1}{c}{{.8555}} &
\multicolumn{1}{c}{{.8572}} \\
\hline
\multicolumn{1}{c|}{\multirow{1}{*}{$l_\mathrm{w}$ hops}}
& \multicolumn{1}{c}{{0}}
& \multicolumn{1}{c}{{1}} & \multicolumn{1}{c}{{2}} &
\multicolumn{1}{c}{{3}} \\ \hline \multicolumn{1}{c|}{ML-1M} & 
\multicolumn{1}{c}{{n/a}} &  
\multicolumn{1}{c}{{.9261}} & \multicolumn{1}{c}{{\textbf{.9267}}} & \multicolumn{1}{c}{{.9262}}  \\ 
\multicolumn{1}{c|}{LFM-1b} & 
\multicolumn{1}{c}{{n/a}} & 
\multicolumn{1}{c}{{.9438}} & \multicolumn{1}{c}{{.9445}} &
\multicolumn{1}{c}{{\textbf{.9447}}}  \\
\multicolumn{1}{c|}{AZ-book} & 
\multicolumn{1}{c}{{n/a}} & 
\multicolumn{1}{c}{{.8568}} & \multicolumn{1}{c}{{.8611}} &
\multicolumn{1}{c}{{\textbf{.8618}}}  \\
\hline
\multicolumn{1}{c|}{\multirow{1}{*}{$l_\mathrm{d}$ hops}} & \multicolumn{1}{c}{{0}} &  \multicolumn{1}{c}{{1}} & \multicolumn{1}{c}{{2}} &
\multicolumn{1}{c}{{3}} \\ \hline \multicolumn{1}{c|}{ML-1M} & 
\multicolumn{1}{c}{{n/a}} & 
\multicolumn{1}{c}{{.9261}} & \multicolumn{1}{c}{{\textbf{.9269}}} & \multicolumn{1}{c}{{.9250}}  \\ 
\multicolumn{1}{c|}{LFM-1b} & 
\multicolumn{1}{c}{{n/a}} & 
\multicolumn{1}{c}{{.9438}} & \multicolumn{1}{c}{{\textbf{.9441}}} & \multicolumn{1}{c}{{.9440}}  \\
\multicolumn{1}{c|}{AZ-book} & 
\multicolumn{1}{c}{{n/a}} & 
\multicolumn{1}{c}{{.8552}} & \multicolumn{1}{c}{{\textbf{.8621}}} & \multicolumn{1}{c}{{.8613}} \\
\hline
\end{tabular}
\label{tb:KGPI_by_P_W_D_hops}
\end{table}

\begin{table}[t]
\small
\centering
\caption{$\mathit{AUC}$ of \system with different embedding size $s$.}
\vspace{-0.5 pc}
\scalebox{1.0}{
\begin{tabular}{p{2.5cm}|ccccccc}
\hline \multicolumn{1}{c|}{\multirow{1}{*}{$s$}} &
\multicolumn{1}{c}{{4}} & \multicolumn{1}{c}{{8}}& \multicolumn{1}{c}{{16}} &
\multicolumn{1}{c}{{32}} & \multicolumn{1}{c}{{64}} & \multicolumn{1}{c}{{128}} \\ \hline \multicolumn{1}{c|}{ML-1M} 
& \multicolumn{1}{c}{{.9037}} 
& \multicolumn{1}{c}{{.9217}} 
& \multicolumn{1}{c}{{.9259}} 
& \multicolumn{1}{c}{{\textbf{.9279}}} 
& \multicolumn{1}{c}{{.9250}}
& \multicolumn{1}{c}{{.9247}}\\ 
\multicolumn{1}{c|}{LFM-1b}
& \multicolumn{1}{c}{{.9247}} 
& \multicolumn{1}{c}{{.9468}} 
& \multicolumn{1}{c}{{.9538}} 
& \multicolumn{1}{c}{{\textbf{.9574}}} 
& \multicolumn{1}{c}{{{.9562}}}
& \multicolumn{1}{c}{{.9538}}\\
\multicolumn{1}{c|}{AZ-book} 
& \multicolumn{1}{c}{{.8353}}
& \multicolumn{1}{c}{{.8471}}
& \multicolumn{1}{c}{{.8616}} 
& \multicolumn{1}{c}{{\textbf{.8664}}}
& \multicolumn{1}{c}{{.8598}}
& \multicolumn{1}{c}{{.8539}}\\
\hline 
\end{tabular}}
\label{tb:KGPH_by_dimension}
\end{table}

\subsection{Parameter Sensitivity}

Below, we investigate the parameter sensitivity in \system.




\paragraph{Preference set sample size $K_m$} 
Table~\ref{tb:KGPH_by_me_preference} shows that the performance of \system
improves when $K_m$ is set to a larger value, with the exception of AZ-book. \system
achieves the best performance on AZ-book when $K_m$ is set to 32, which we
attribute to its low number of user-interacted items, as shown in
Table~\ref{tb:datasetstatistic}. That is, when there are few
user-interacted items, a small $K_m$ still allows \system to find enough information
to represent the user.


\paragraph{Neighborhood entity sample size $K_n$}
The influence of the size of neighborhood nodes
is shown in Table~\ref{tb:KGPH_by_me_preference}. \system achieves the
best performance when this is set to 16 or 32, perhaps due to the noise introduced
when $K_n$ is too large. 

\paragraph{Number of preference hops $l_\mathrm{p}$} 
The impact of $l_\mathrm{p}$ is shown in
Table~\ref{tb:KGPI_by_P_W_D_hops}. We conducted experiments with
$l_\mathrm{p}$ set to 0, that is, we only use user-clicked items
$h^{1}_{i}$ to calculate user representation. The results show that when
$l_\mathrm{p}$ hop is set to 1, \system achieves the best performance, whereas
again larger values of $l_\mathrm{p}$ result in less relevant entities and
thus more noise, consistent with~\cite{DBLP:journals/corr/abs-1803-03467}.
 
 
\paragraph{Number of wide hops $l_\mathrm{w}$ and deep hops $l_\mathrm{d}$} 
Table~\ref{tb:KGPI_by_P_W_D_hops} shows the effect of varying the number 
of the wide hops $l_\mathrm{w}$ and deep hops $l_\mathrm{d}$.
\system achieves better
performance when the number of hops is set to 2 over 1, suggesting that increasing the
hops enables the modeling of high-order connectivity and hence 
enhances the performance. However, the performance drops when the number of
hops becomes even larger, i.e., 3, suggesting that considering second-order
relations among entities is sufficient, consistent
with~\cite{2019arXiv191008288S,DBLP:journals/corr/abs-1904-12575}.


\paragraph{Dimension of embedding size $s$} The results when varying the embedding size
are shown in Table~\ref{tb:KGPH_by_dimension}. Increasing $s$
initially boosts the performance as a larger $s$ contains more useful
information of users and entities, whereas setting $s$ too large leads to
overfitting.

\section{Conclusion}
We propose \system, a GNN-based recommendation model which improves
representations of items from both the user view and the entity view. Given
both user- and entity-view features, \system gathers personalized
knowledge information in the KG (user view) and further considers the difference
among layers (entity view) to ultimately enhance item representations. 
Extensive experiments show the superiority of \system.
In addition, the ablation experiment verifies the effectiveness of each
proposed component. 

As the proposed components are general, the method could also be applied to leverage
structural information such as social networks or item contexts in the
form of knowledge graphs. We believe \system can be widely used in related
applications.

\bibliographystyle{ACM-Reference-Format}
\bibliography{sample-base}

\end{document}